\documentclass[11pt]{article}

\usepackage{hyperref,amsfonts,amsmath,amssymb,graphicx,bm,subfigure,a4wide,cite}

\begin{document}

\title{\textbf{Neutrino spin and flavor oscillations in gravitational fields}}

\author{Maxim Dvornikov\thanks{maxdvo@izmiran.ru}
\\
\small{\ Pushkov Institute of Terrestrial Magnetism, Ionosphere} \\
\small{and Radiowave Propagation (IZMIRAN),} \\
\small{108840 Moscow, Troitsk, Russia}}

\date{}

\maketitle

\begin{abstract}
We study spin and flavor oscillations of astrophysical neutrinos under the influence of external fields in curved spacetime. First, we consider spin oscillations in case of neutrinos gravitationally scattered off a rotating supermassive black hole surrounded by a thin magnetized accretion disk. We find that the gravitational interaction only does not result in the spin-flip of scattered ultrarelativistic neutrinos. Realistic magnetic fields lead to the significant reduction of the observed flux of neutrinos possessing reasonable magnetic moments. Second, we study neutrino flavor oscillations in stochastic gravitational waves (GWs). We derive the effective Hamiltonian for neutrinos interacting with a plane GW having an arbitrary polarization. Then, we consider stochastic GWs with arbitrary correlators of amplitudes. The equation for the density matrix for neutrino oscillations is solved analytically and the probabilities to detect certain neutrino flavors are derived. We find that the interaction of neutrinos, emitted by a core-collapsing supernova, with the stochastic GW background results in the several percent change of the neutrino fluxes. The observability of the predicted effects is discussed.
\end{abstract}

The experimental confirmation of neutrino oscillations is an unambiguous indication on the validity of certain extensions of the standard model in which neutrinos are massive and mixed particles. External fields, such as the electroweak interaction with matter and electromagnetic fields, are known to modify the process of neutrino oscillations. Gravity, in spite of its weakness in comparison with other interactions, can also significantly affect neutrino oscillations. In the present work, we are interested in the neutrino propagation and oscillations in various gravitational backgrounds, like a rotating black hole (BH) and gravitational waves (GWs). The subject of the present research is inspired by a significant experimental success in the study of these gravitational fields (see, e.g., Refs.~\cite{Abb21,Aki22}).

There are two main types of neutrino oscillations. First, we mention neutrino spin oscillations, in which left polarized active neutrinos are converted to right polarized sterile particles. Second, there are neutrino flavor oscillations, where transitions between different neutrino generations are possible. Previously, we considered both spin and flavor oscillations in gravitational fields in Refs.~\cite{Dvo06,Dvo13,Dvo19b,Dvo20a,Dvo19,Dvo20,Dvo21,Dvo20b,Dvo21b}. In the present work, we summarize our achievements in the studies of neutrino oscillations in a curved spacetime.


First, we consider ultrarelativistic neutrinos which are scattered by a rotating BH surrounded by a thin magnetized accretion disk. A neutrino is supposed to be a Dirac particle possessing a nonzero magnetic moment. Using the Boyer-Lindquist coordinates $x^{\mu}=(t,r,\theta,\phi)$, the spacetime of a rotating BH has the following metric:
\begin{equation}\label{eq:Kerrmetr}
  \mathrm{d}s^{2} =
  \left(
    1-\frac{rr_{g}}{\Sigma}
  \right)
  \mathrm{d}t^{2}+2\frac{rr_{g}a\sin^{2}\theta}{\Sigma}\mathrm{d}t\mathrm{d}\phi-\frac{\Sigma}{\Delta}\mathrm{d}r^{2}-
  \Sigma\mathrm{d}\theta^{2}-\frac{\Xi}{\Sigma}\sin^{2}\theta\mathrm{d}\phi^{2},
\end{equation}
where $\Delta=r^{2}-rr_{g}+a^{2}$, $\Sigma=r^{2}+a^{2}\cos^{2}\theta$, and $\Xi=
  \left(
    r^{2}+a^{2}
  \right)
  \Sigma+rr_{g}a^{2}\sin^{2}\theta$.
Here we take that the gravitational constant equals to one. Thus, we have that the angular momentum of BH
is $J=Ma$ and its mass is $M=r_{g}/2$, where $r_{g}$ is the Schwarzschild radius.

The detailed description of the motion of an ultrarelativistic test particle in the Kerr metric in Eq.~\eqref{eq:Kerrmetr} is given in Ref.~\cite{Cha83}. We just recall that the trajectory and the law of motion can be found in quadratures. Besides the conserved energy $E$, any trajectory can be labeled by the angular momentum $L$ and the Carter constant $Q$, which is positive if we study the scattering problem.

The quasiclassical description of the neutrino spin is based on the formalism developed in Refs.~\cite{Dvo06,PomKhr98}, where a particle spin is supposed to be parallel transported along geodesics. It is convenient to follow the neutrino spin dynamics in the locally Minkowskian frame
$x_{a}=e_{a}^{\ \mu}x_{\mu}$. The explicit form of the vierbein vectors $e_{a}^{\ \mu}$, $a=0,\dots,3$, is given in Ref.~\cite{Dvo13}. They satisfy the relation $e_{a}^{\ \mu}e_{b}^{\ \nu}g_{\mu\nu}=\eta_{ab}=\text{diag}(1,-1,-1,-1)$.

The evolution of the invariant three vector of the neutrino spin $\bm{\zeta}$ obeys the equation
$
  \dot{\bm{\bm{\zeta}}}=2(\bm{\bm{\zeta}}\times\bm{\bm{\Omega}})
$, 
where
\begin{equation}\label{eq:vectG}
  \bm{\bm{\Omega}}=\frac{1}{U^{t}}
  \left\{
    \frac{1}{2}
    \left[
      \mathbf{b}_{g}+\frac{1}{1+u^{0}}
      \left(
        \mathbf{e}_{g}\times\mathbf{u}
      \right)
    \right]+
    \mu
    \left[
      u^{0}\mathbf{b}-\frac{\mathbf{u}(\mathbf{u}\mathbf{b})}{1+u^{0}}+(\mathbf{e}\times\mathbf{u})
    \right]
  \right\}.
\end{equation}
Here $u^{a}=(u^{0},\mathbf{u})$
is the four velocity, $U^t$ is the component of the four velocity in world coordinates, $G_{ab}=(\mathbf{e}_{g},\mathbf{b}_{g})=\gamma_{abc}u^{c}$, $\gamma_{abc}$ are the Ricci rotation coefficients, $f_{ab}=(\mathbf{e},\mathbf{b})$
is the electromagnetic field tensor, and $\mu$ is the diagonal neutrino magnetic moment. All quantities labeled with Latin indexes $a,b,\dotsc$ are given in the locally Minkowskian frame.

We consider the neutrino scattering off supermassive BH (SMBH) with the mass $M=10^{8}M_{\odot}$. The incoming flux of neutrinos is taken to be parallel to the equatorial plane of BH. We neglect the electroweak interaction of neutrinos with plasma of the accretion disk since the disk is supposed to be slim. The distribution of the magnetic field around a rotating BH is studied in Ref.~\cite{Wal74}. Additionally, we take that the amplitude of the magnetic field scales with the distance to the BH center as $B\propto B_{0}r^{-5/4}$~\cite{BlaPay82}. It guarantees that the magnetic field vanishes at the accretion disk edge. Here, $B_{0}=3.2\times10^{2}\,\text{G}$
(for $M=10^{8}M_{\odot}$) is the strength of the magnetic field in the vicinity
of BH at $r\sim r_{g}$~\cite{Bes10}. Moreover, we take into account only the poloidal magnetic field since the accretion disk is assumed to be thin. The value of the neutrino magnetic moment is in the range $\mu=(10^{-14}-10^{-13})\mu_{\mathrm{B}}$, where $\mu_{\mathrm{B}}$
is the Bohr magneton. It is consistent with the model independent constraint on the Dirac
neutrino magnetic moment in Ref.~\cite{Bel05} and with the astrophysical
upper bound on the neutrino magnetic moment in Ref.~\cite{Via13}.

We suppose that incoming neutrinos are left polarized, that is valid for ultrarelativistic particles in frames of the standard model. If the neutrino polarization changes in the process of the interaction with external fields in the gravitational scattering, we observe the reduced flux $F_{\nu}=P_{\mathrm{LL}}F_{0}$, where $P_{\mathrm{LL}}$ is the survival probability of neutrino spin oscillations, which is found in solving the spin evolution equation along each neutrino trajectory, and $F_{0}$ is the flux of scalar particles. We shall study the ratio $F_{\nu}/F_{0}$ for neutrinos gravitationally scattered off a Kerr SMBH taking into account the interaction of the neutrino
magnetic moment with a magnetic field in an accretion disk.

\begin{figure}
  \centering
  \subfigure[]
  {\label{fig:grava}
  \includegraphics[scale=.35]{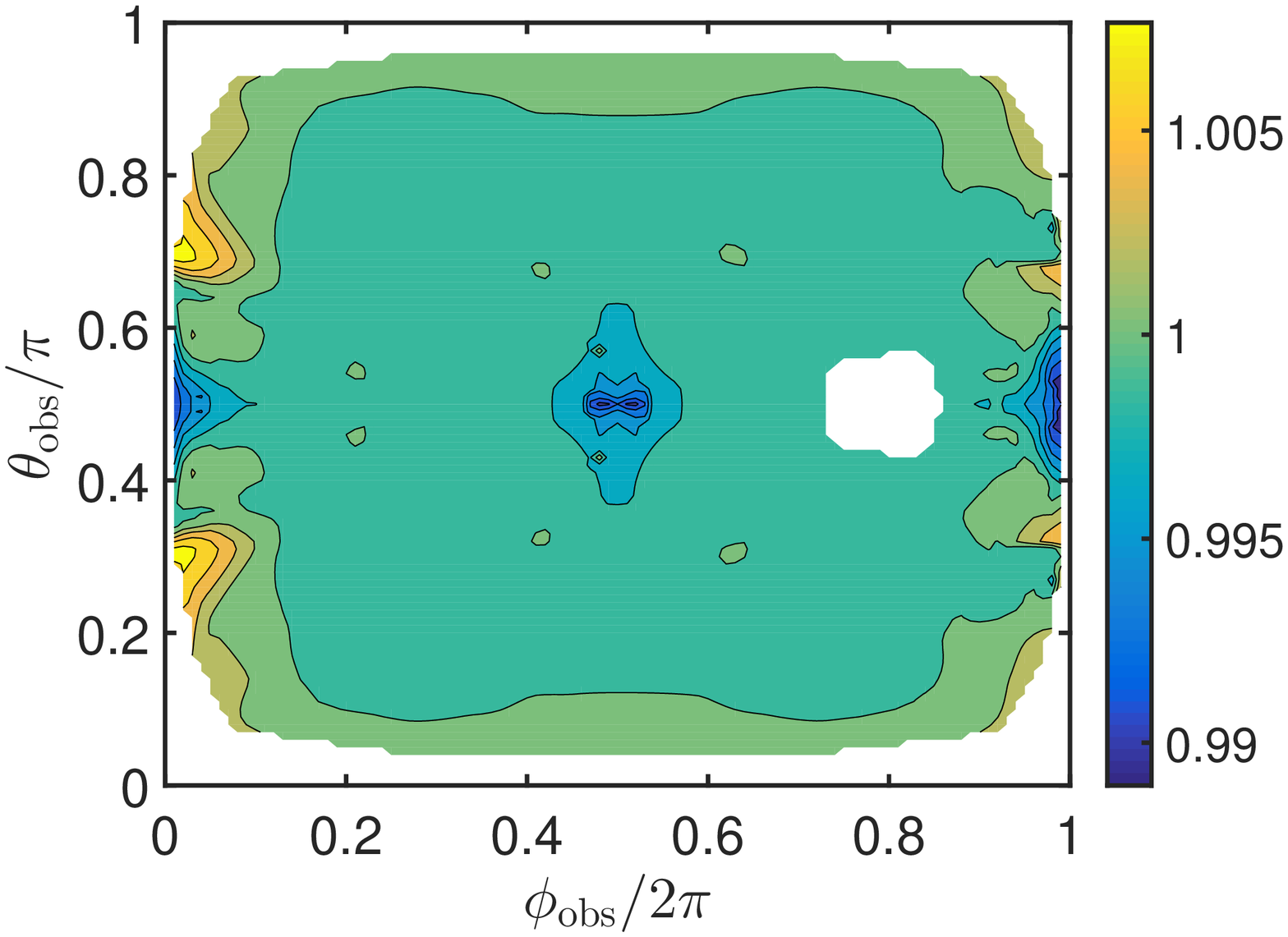}}
  \hskip-.3cm
  \subfigure[]
  {\label{fig:gravb}
  \includegraphics[scale=.35]{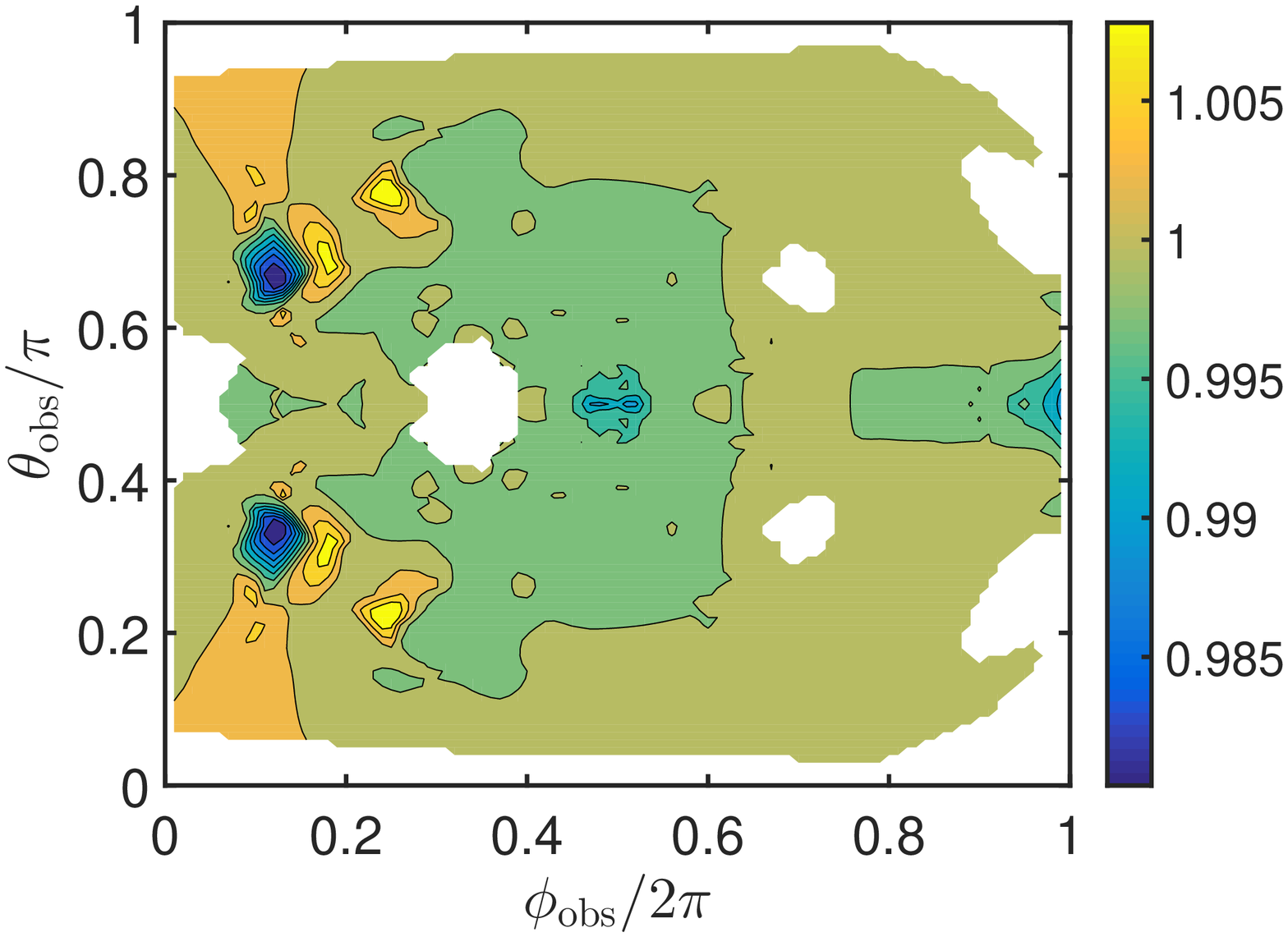}}
  \protect
\caption{The ratio $F_{\nu}/F_{0}$ for neutrinos gravitationally scattered off BHs with different
angular momenta. (a) $a=2\times10^{-2}M$; (b) $a=0.98M$.\label{fig:grav}}
\end{figure}

First, we study the impact of gravity only on the spin evolution in the scattering of ultrarelativistic neutrinos. The neutrino flux in shown in Fig.~\ref{fig:grav} for different spins of BH. We can see that $F_\nu = F_0$ with very high accuracy. It means that gravity does not lead to a spin-flip of ultreraletivistic neutrinos.

\begin{figure}
  \centering
  \subfigure[]
    {\label{fig:magna}
    \includegraphics[scale=.35]{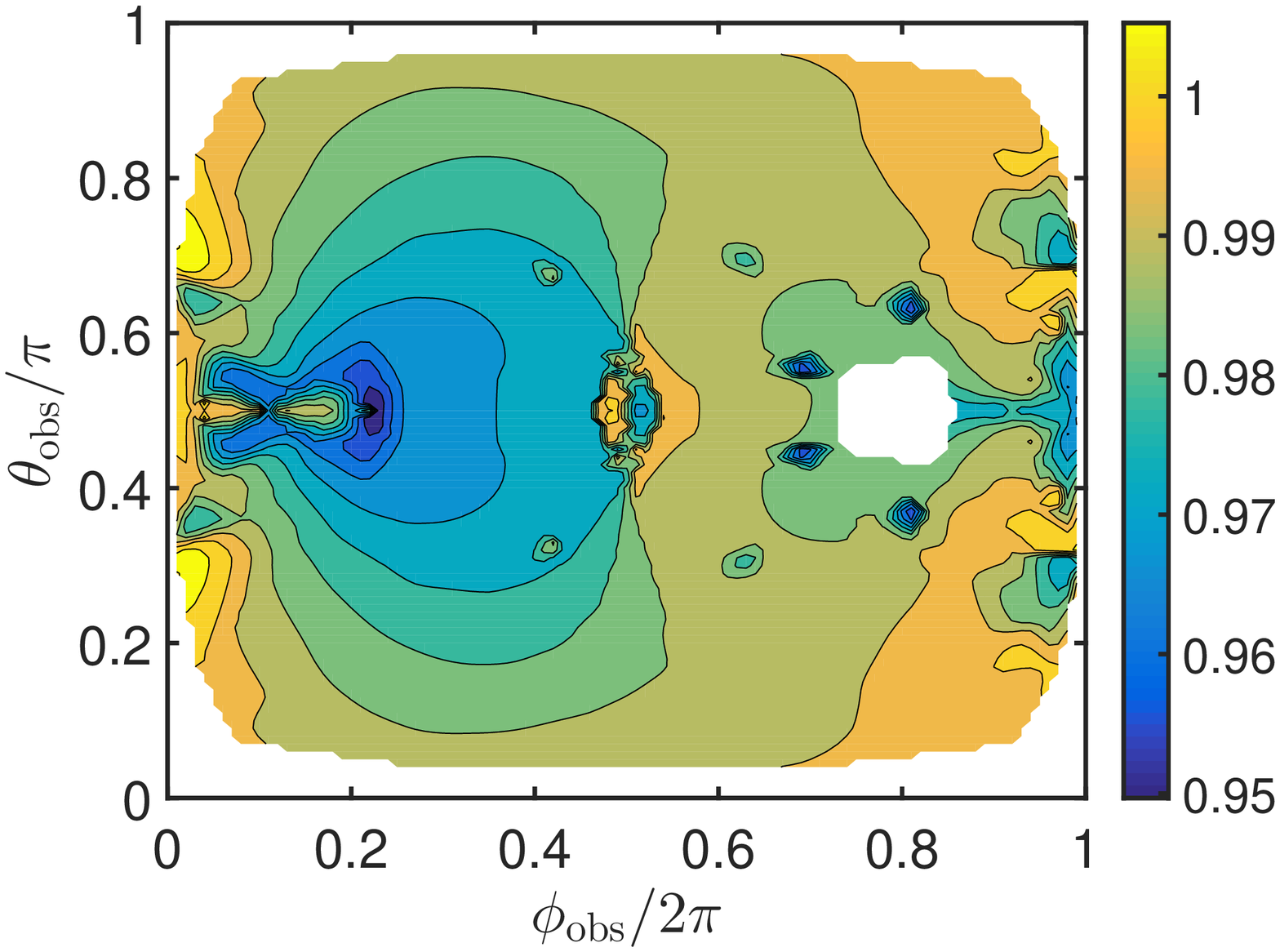}}
  \hskip-.3cm
  \subfigure[]
    {\label{fig:magnb}
    \includegraphics[scale=.35]{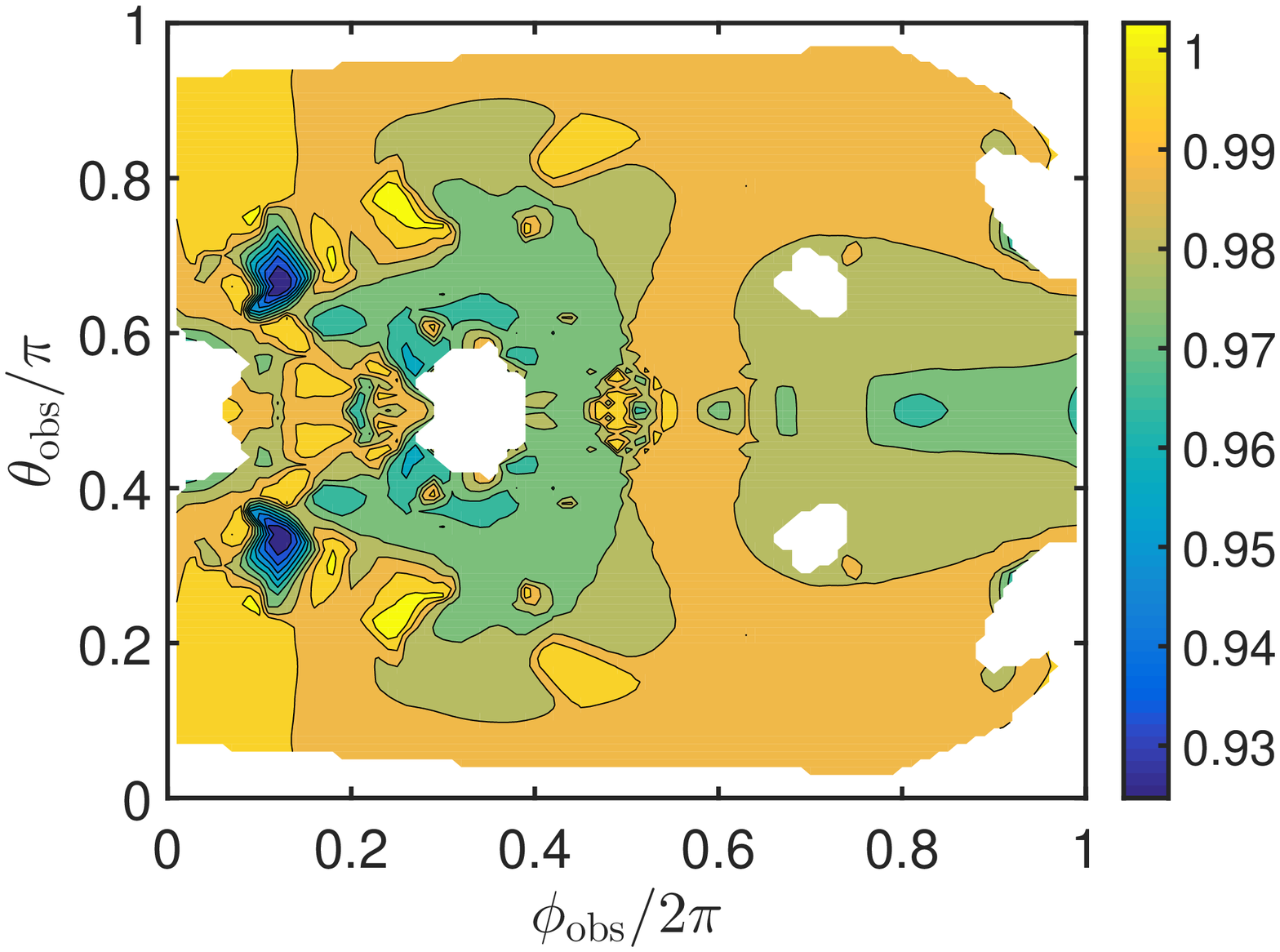}}
  \\
  \subfigure[]
    {\label{fig:magnc}
    \includegraphics[scale=.35]{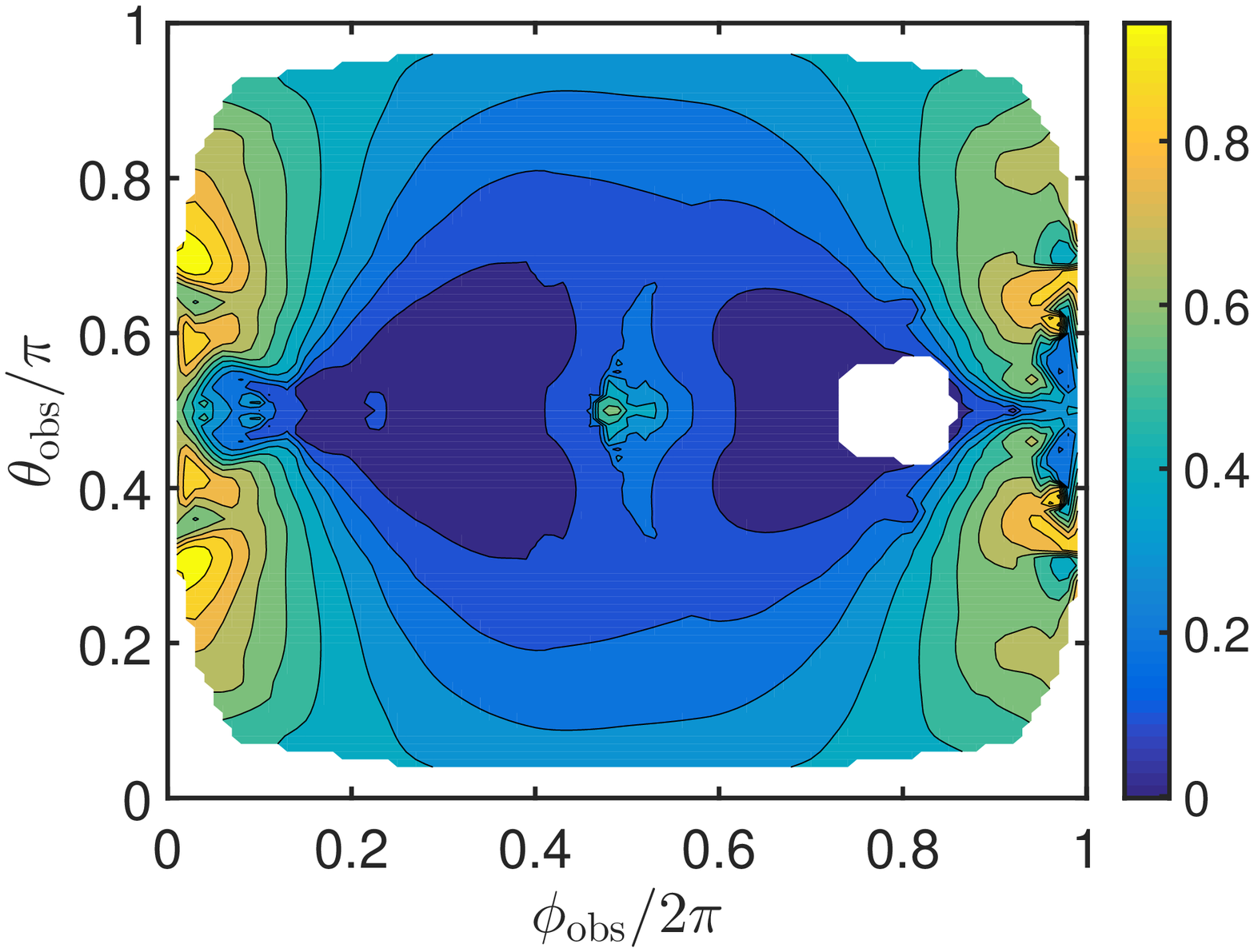}}
  \hskip-.3cm
  \subfigure[]
    {\label{fig:magnd}
    \includegraphics[scale=.35]{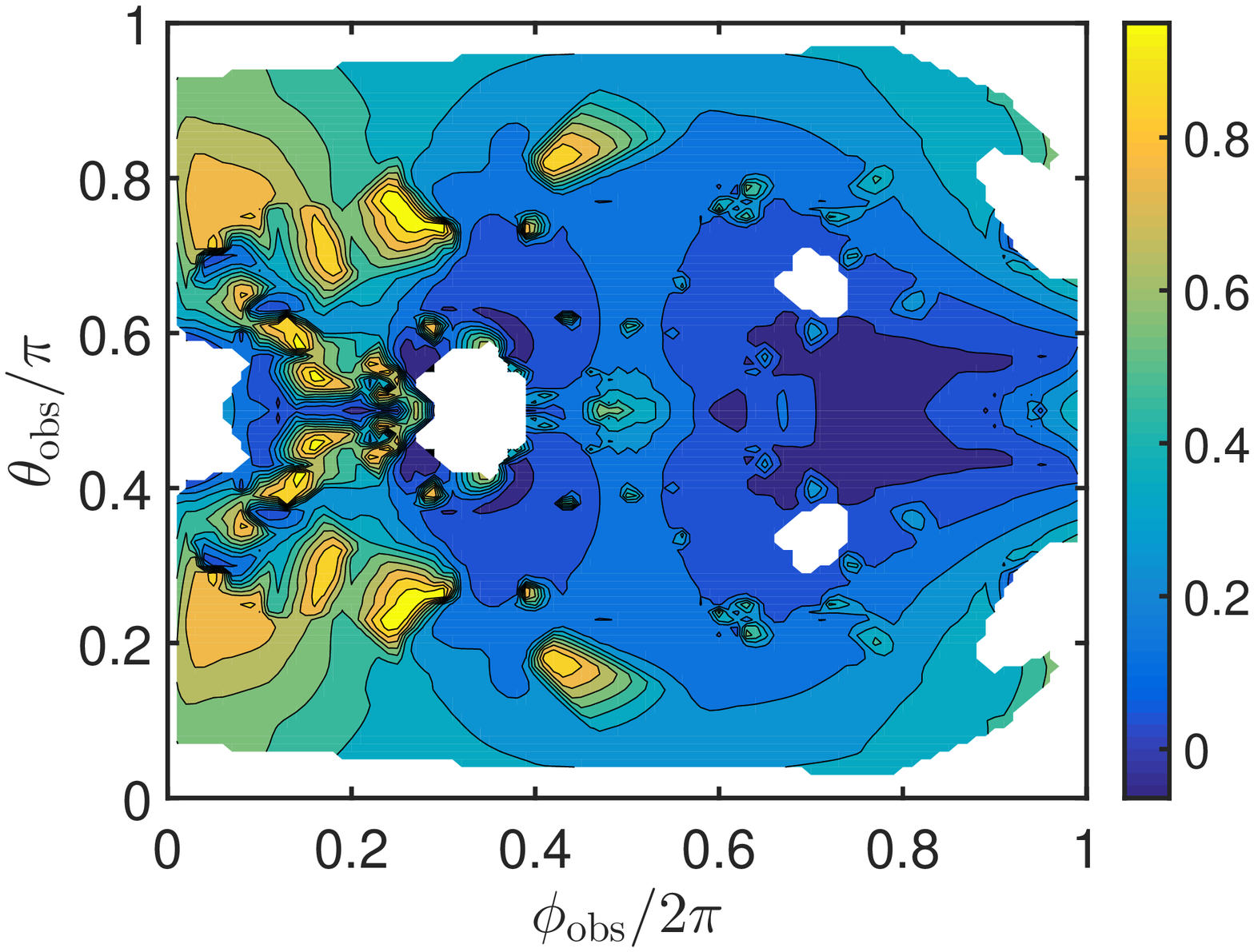}}
  \protect
\caption{The same as in Fig.~\ref{fig:grav} accounting for the neutrino magnetic
interaction. Panels (a) and (b): $\mu=10^{-14}\mu_{\mathrm{B}}$;
panels (c) and (d): $\mu=10^{-13}\mu_{\mathrm{B}}$. Panels (a)
and (c): $a=2\times10^{-2}M$; Panels (b) and (d):
$a=0.98M$.\label{fig:magn}}
\end{figure}

We can see that a nonzero magnetic interaction is necessary to produce a neutrino spin-flip. We show $F_{\nu}/F_{0}$ for different $V_{\mathrm{B}}=\mu B_{0}r_{g}$ and spins of BH in Fig.~\ref{fig:magn}. The flux of neutrinos is reduced by up to $(5\pm1)\%$ for $\mu=10^{-14}\mu_{\mathrm{B}}$ in Figs.~\ref{fig:magna} and~\ref{fig:magnb}, which is consistent with the results of Ref.~\cite{Dvo21}. If we take $\mu=10^{-13}\mu_{\mathrm{B}}$, the neutrino flux becomes almost vanishing for certain scattering angles $\phi_\mathrm{obs}$ and $\theta_\mathrm{obs}$; cf. Figs.~\ref{fig:magnc} and~\ref{fig:magnd}.

Now, we turn to the consideration of neutrino flavor oscillations in stochastic GWs. The system of three active flavor neutrinos $\nu_{\lambda}$, $\lambda=e,\mu,\tau$,
with the nonzero mixing, as well as under the influence of a plane GW
with an arbitrary polarization, obeys the following Schr\"odinger equation:
$
  \mathrm{i}\dot{\nu}=(H_{0}+H_{1})\nu
$, 
where $\nu^\mathrm{T} = (\nu_e,\nu_\mu,\nu_\tau)$, $H_{0}=UH_{m}^{(\mathrm{vac})}U^{\dagger}$ is the effective
Hamiltonian for vacuum oscillations in the flavor eigenstates basis,
$H_{m}^{(\mathrm{vac})}=\tfrac{1}{2E}\text{diag}\left(0,\Delta m_{21}^{2}, \Delta m_{31}^{2}\right)$
is the vacuum effective Hamiltonian for the mass eigenstates $\psi_{a}$,
$a=1,2,3$, and $U$ is the unitary matrix which
relates flavor and mass bases: $\nu_{\lambda}=U_{\lambda a}\psi_{a}$.\sloppy

The Hamiltonian $H_{1}$, which describes
the neutrino interaction with GW, has the form $H_{1}=UH_{m}^{(g)}U^{\dagger}$~\cite{Dvo21b},
where $H_{m}^{(g)} = H_{m}^{(\mathrm{vac})}\left(A_{c}h_{+}+A_{s}h_{\times}\right)$, and
\begin{equation}\label{eq:Hgmass}
  A_{c} = \frac{1}{2}\sin^{2}\vartheta\cos2\varphi\cos\Phi,
  \quad
  A_{s} = \frac{1}{2}\sin^{2}\vartheta\sin2\varphi \sin\Phi.
\end{equation}
Here,
$h_{+,\times}$ are the amplitudes corresponding to `plus' and `times'
polarizations of GW, $\Phi = \omega t(1-\cos\vartheta)$, $\omega$ is the frequency of GW, $\vartheta$
and $\varphi$ are the spherical angles fixing the neutrino momentum
with respect to the wave vector of GW, which is supposed to propagate
along the $z$-axis.

Now we consider the situation when a neutrino interacts with stochastic
GWs. In this case, the angles $\vartheta$ and $\varphi$, as well
as the amplitudes $h_{+,\times}$, are random functions of time. To
study the neutrino motion in such a background, it is more convenient
to deal with the density matrix $\rho$. We can also introduce the density matrix in the interaction picture, $\rho_{\mathrm{int}}=\exp(\mathrm{i}H_{0}t)\rho\exp(-\mathrm{i}H_{0}t)$. Assuming that the background of GWs is Gaussian and appropriately averaging the evolution equation for $\rho_{\mathrm{int}}$, we get that $\tfrac{\mathrm{d}}{\mathrm{d}t}
  \left\langle
    \rho_{\mathrm{int}}
  \right\rangle (t)=
  -g(t)\tfrac{3}{64}[H_{0},[H_{0},
  \left\langle
    \rho_{\mathrm{int}}
  \right\rangle (t)]]$, where
\begin{equation}\label{eq:rhoIeqfin}
  g(t) = \frac{1}{2}\int_{0}^{t}\mathrm{d}t_{1}
  \left(
    \left\langle
      h_{+}(t)h_{+}(t_{1})
    \right\rangle +
    \left\langle
      h_{\times}(t)h_{\times}(t_{1})
    \right\rangle
  \right).
\end{equation}
Here the correlators of the amplitudes $\left\langle h_{+,\times}(t)h_{+,\times}(t_{1})\right\rangle$ are supposed to be arbitrary. Using the function $\Omega(f)=\tfrac{f}{\rho_{c}}\tfrac{\mathrm{d}\rho_{\mathrm{GW}}}{\mathrm{d}f}$~\cite{Chr19},
where $\rho_{\mathrm{GW}}$ is the energy density of GW and $\rho_{c}=0.53\times10^{-5}\,\text{Gev}\cdot\text{cm}^{-3}$
is the closure energy density of the universe, we can rewrite $g(t)$ in Eq.~\eqref{eq:rhoIeqfin} in the form,
$ 
  g(t)=\tfrac{4G\rho_{c}}{\pi^{2}}\smallint_{0}^{\infty}\tfrac{\mathrm{d}f}{f^{4}}\sin(2\pi ft)\Omega(f)
$. 
Here $G=6.9\times10^{-39}\,\text{GeV}^{-2}$ is the Newton's constant and $f$ is the frequency measured in Hz. 

We can find the analytical solution for the density matrix and write down the probability to observe a certain neutrino flavor. If we study the propagation of supernova (SN) neutrinos, which are characterized by the ratio of the initial fluxes $\left(F_{\nu_{e}}:F_{\nu_{\mu}}:F_{\nu_{\tau}}\right)_{\mathrm{S}}=(1:0:0)$, the deviation of the probability from the vacuum oscillations value, $\Delta P_{\lambda}=P_{\lambda}^{(g)}-P_{\lambda}^{(\mathrm{vac})}$, is
\begin{align}\label{eq:DeltaP21}
  \Delta P_{\lambda}(x)= & 2
  \left[
    1-\exp
    \left(
      -\Gamma
    \right)
  \right]
  \bigg[
    \text{Re}
    \left[
      U_{\lambda2}U_{\lambda1}^{*}U_{e2}^{*}U_{e1}
    \right]
    \cos
    \left(
      2\pi\frac{x}{L_{21}}
    \right)
    \nonumber
    \\
    & +
    \text{Im}
    \left[
      U_{\lambda2}U_{\lambda1}^{*}U_{e2}^{*}U_{e1}
    \right]
    \sin
    \left(
      2\pi\frac{x}{L_{21}}
    \right)
  \bigg],
  \quad
  \Gamma= \frac{3\pi^{2}}{16 L_{21}^{2}}\int_{0}^{x}g(t)\mathrm{d}t,
\end{align}
where $L_{21}=\tfrac{4\pi E}{\Delta m_{21}^{2}}$ is the oscillations length for the solar neutrino oscillations channel and $x\approx t$ is the distance traveled by the neutrino beam.

Now we should fix the source of the GW background. We suppose that stochastic GWs are emitted by merging SMBHs. In this case~\cite{Ros11}, $\Omega(f) = \Omega_{0}\sim10^{-9}$ if $f_{\mathrm{min}}<f<f_{\mathrm{max}}$, where $f_{\mathrm{min}}\sim10^{-10}\,\text{Hz}$
and $f_{\mathrm{max}}\sim10^{-1}\,\text{Hz}$, and $\Omega(f) = 0$ otherwise.

\begin{figure}
  \centering
  \subfigure[]
  {\label{fig:deltaFa}
  \includegraphics[scale=.35]{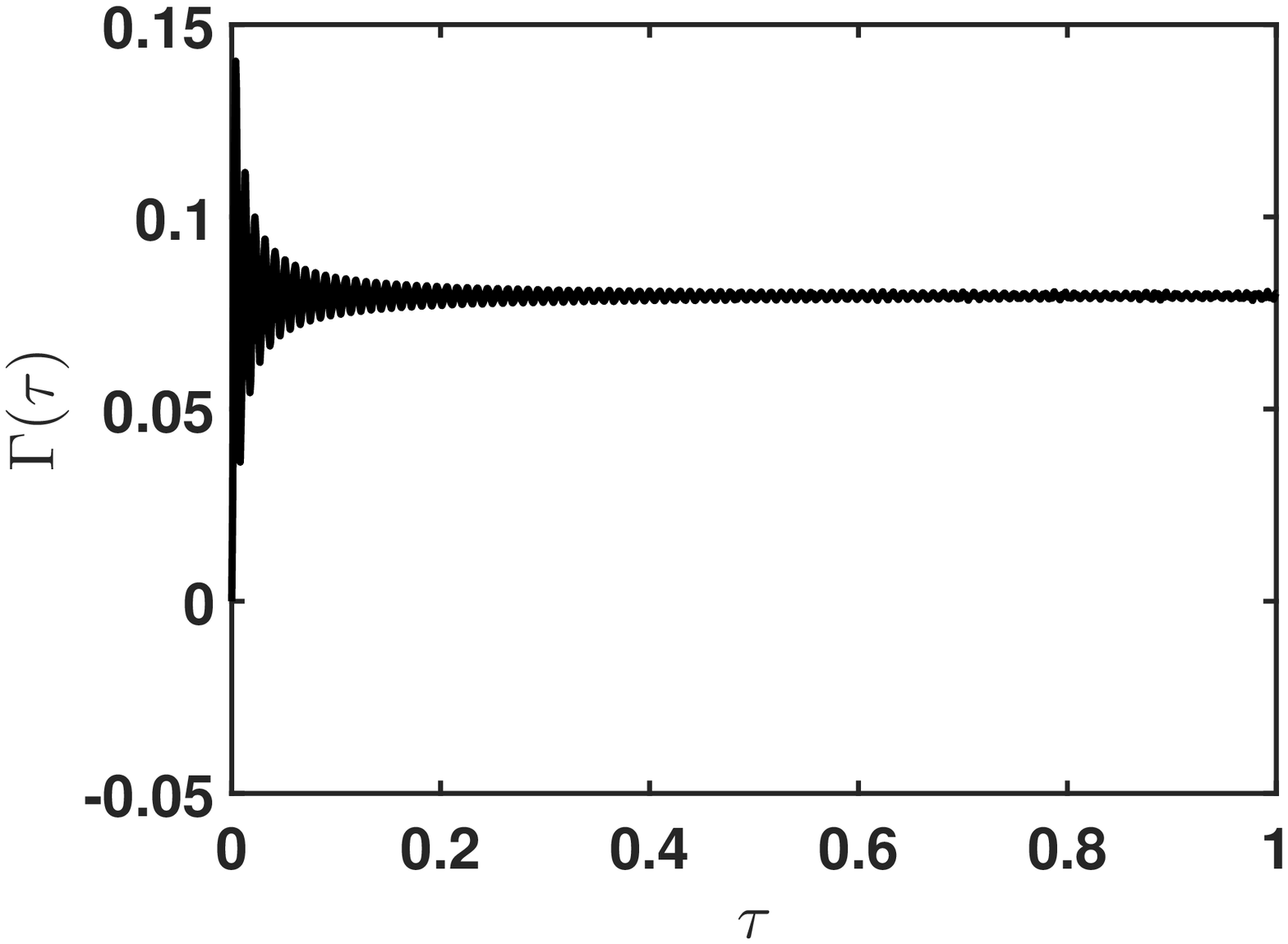}}
  \hskip-.6cm
  \subfigure[]
  {\label{fig:deltaFb}
  \includegraphics[scale=.35]{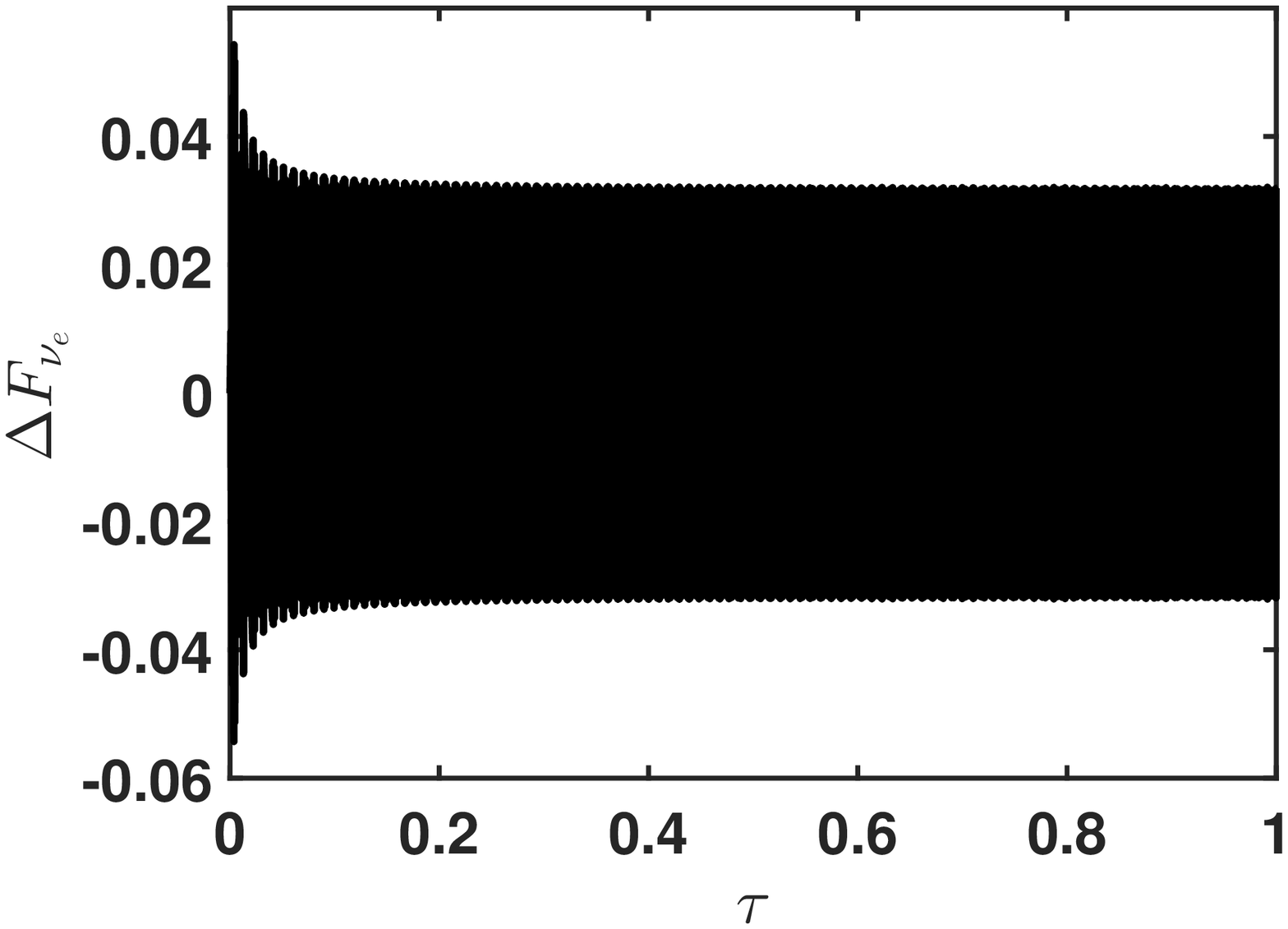}}
  \protect
  \caption{(a) The parameter $\Gamma$ versus the neutrino beam propagation length
  $x=\tau L$; (b) The correction the flux of $\nu_e$, $\Delta F_{\nu_{e}}\propto\Delta P_{\nu_e}$,
  owing to the neutrino interaction with stochastic GWs. The parameters
  of the system are $\Delta m_{21}^{2}=7.5\times10^{-5}\,\text{eV}^{2}$,
  $E=10\,\text{MeV}$ is the typical SN neutrino energy,  
  $L=10\,\text{kpc}$, $\Omega_{0}=10^{-9}$, and $f_{\mathrm{min}}=10^{-10}\,\text{Hz}$.
  The normal neutrino mass ordering is adopted.\label{fig:deltaF}}
\end{figure}

In Fig.~\ref{fig:deltaFa}, we show the dependence of $\Gamma$ in Eq.~\eqref{eq:DeltaP21} versus the distance passed by neutrinos $\tau = x/L$, where $L = 10\,\text{kpc}$ is the typical Galaxy size. We can see that, at $\tau \to 1$, $\Gamma \to \Gamma_{\oplus}= 8\times10^{-2}$.  In Fig.~\ref{fig:deltaFb}, we show the deviation of the flux of $\nu_e$ from the vacuum oscillations value owing to the interaction with stochastic GWs. We can see that the interaction with GWs results in the $\sim 3\%$ change of the observed flux.

In the present work, we have studied neutrino oscillations in gravitational fields. First, we have considered spin oscillations of neutrinos in their gravitational scattering off rotating BH. We have started with the formulation of the main equations for the neutrino spin under the influence of external fields in curved spacetime. It is defined in the locally Minkowskian frame. Then, we have applied this result to the neutrino spin evolution in the gravitational field of a rotating SMBH surrounded by a thin magnetized accretion disk. We have supposed that the incoming flux of ultrarelativistic neutrinos is parallel to the equatorial plane of BH. However, it is both above and below this plane. Thus, we have generalized our previous results in Refs.~\cite{Dvo20,Dvo21}, where only the equatorial neutrino motion is studied. The neutrino electroweak interaction with plasma of the disk has not been taken into account since the disk is supposed to be thin. Only the poloidal component of the magnetic field in the disk was considered. The strength of the magnetic field is typical for SMBH.

We have obtained that only the gravitational interaction does not result in the spin-flip of ultrarelativistic neutrinos in their scattering off BH. This our finding is in agreement with the results of Ref.~\cite{Lam05}. Thus, the variation of the observed neutrino flux, which depends on the survival probability of neutrino oscillations, is possible only if a magnetic field is present in the system besides gravity. We have found that the neutrino flux can be significantly reduced if the neutrino magnetic moment has realistic values consistent with theoretical and astrophysical constraints.

Then, we have studied neutrino flavor oscillations under the influence of stochastic GWs. We have obtained the effective Hamiltonian for flavor oscillations for massive neutrinos interacting with a plane GW~\cite{Dvo19}. We have also accounted for two independent polarizations of GW~\cite{Dvo21b}. The obtained Hamiltonian was used to describe neutrino flavor oscillations in stochastic GWs with arbitrary correlators of amplitudes~\cite{Dvo20b}.

These findings were applied to analyze the impact of stochastic GWs on the flavor content of neutrinos emitted by a core-collapsing SN. In this situation, the source of neutrinos is almost point like. Moreover, we know the flavor content of emitted SN neutrinos. We have obtained that the major contribution to the observed fluxes is for the GW background created by merging SMBHs. The deviation of the fluxes caused by the neutrino interaction with stochastic GWs can be up to several percent. The maximal effect is for the electron neutrinos flux.

Both spin and flavor oscillations phenomena, predicted in our work, can be observed for astrophysical neutrinos using existing and especially future neutrino telescopes (see, e.g., Ref.~\cite{An16}). These facilities have sufficient sensitivity to detect the deviation of neutrino fluxes owing to the neutrino interaction with external fields in curved spacetime.

\end{document}